\documentclass[%
reprint,
 amsmath,amssymb,
 aps,
]{revtex4-2}

\usepackage{graphicx}% Include figure files
\usepackage{dcolumn}% Align table columns on decimal point
\usepackage{bm}% bold math
\usepackage{multirow}% bold math
\usepackage{caption}
\usepackage{blindtext}
\usepackage{appendix}
\usepackage{graphicx}
\usepackage{amsmath}
\usepackage{xcolor}
\usepackage{ulem}

\begin{document}
\title{Understanding the $L$-shell ionization mechanism through osmium atoms bombarded by 4-6 MeV/u fluorine ions} 
%\title{Role of projectile charge states through L-shell x-ray production cross sections} %in ultra-thin osmium target by 4-6 MeV/u fluorine ions}
\author{Soumya Chatterjee$^1$, Sunil Kumar$^2$, Sarvesh Kumar$^3$, M Oswal$^4$, Biraja Mohanty$^5$, D. Mehta$^5$, D. Mitra$^1$, A. M. P. Mendez$^6$, D. M. Mitnik$^6$, C. C. Montanari$^6$, L. Sarkadi$^7$, and T. Nandi$^{8*}$}
%\author{Shashank Singh$^1$, Soumya Chatterjee$^2$, Prashant Sharma$^3$, Mumtaz Oswal$^4$, Sunil Kumar$^1$,  D. Mitra$^1$ and T. Nandi$^{7*}$}
%\affiliation{$^1$Department of Physics, Panjab University, Chandigarh-160014, India.}
\affiliation{$^1$Department of Physics, University of Kalyani, Kalyani, West Bengal-741235, India.}
\affiliation{$^2$Govt. Degree College, Banjar, Kullu (HP) 175123, India.}
\affiliation{$^3$Atomic and Molecular Collisions Laboratory, CEFITEC, Department of Physics.}
\affiliation{$^4$Department of Physics, Dev Samaj College for Women, Sector 45 B, Chandigarh-160014, India.}
\affiliation{$^5$Department of Physics, Panjab University, Chandigarh-160014, India.}
\affiliation{$^6$ Instituto de Astronomía y Física del Espacio, CONICET and Universidad de Buenos Aires, Buenos Aires, Argentina}
\affiliation{$^7$Institute for Nuclear Research of the Hungarian Academy of Sciences (MTA Atomki), H-4001
Debrecen, Pf. 51, Hungary.}
%\affiliation{$^5$Govt. Degree college, Banjar, Kullu, Himachal Pradesh-175123, India.}
%\affiliation{$^5$Department of Particle and Astrophysics, Weizmann Institute of Science, Rehovot 76100, Israel}
\affiliation{$^{8*}$Inter-University Accelerator Centre, Aruna Asaf Ali Marg, Near Vasant Kunj, New Delhi-110067, India.}
\thanks {Email:\hspace{0.0cm} nanditapan@gmail.com. Present address: 1003 Regal, Mapsko Royal Ville, Sector-82, Gurgaon-122004, India.}

\begin{abstract}
The $L$-subshell ionization mechanism is studied in an ultra-thin osmium target bombarded by 4-6 MeV/u fluorine ions. Multiple ionization effects in the collisions are considered through the change of fluorescence and Coster-Kronig yields while determining $L$-subshell ionization cross sections from $L$-line x-ray production cross sections. The $L$-subshell ionization, as well as $L$-shell x-ray production cross sections so obtained, are compared with various theoretical approximations. The Coulomb direct ionization contributions is studied by (i) the relativistic semi-classical approximations (RSCA), (ii) the shellwise local plasma approximation (SLPA), and (iii) the ECUSAR theory, along with the inclusion of the vacancy sharing among the subshells by the coupled-states model (CSM) and the electron capture (EC) by a standard formalism. We find that the ECUSAR-CSM-EC describes the measured excitation function curves the best. However, the theoretical calculations are still about a factor of two smaller than the measured values. Such differences are resolved by re-evaluating the fluorescence and the Coster-Kronig yields. This work demonstrates that, in the present energy range, the heavy-ion induced inner-shell ionization of heavy atoms can be understood by combining the basic mechanisms of the direct Coulomb ionization, the electron capture, the multiple ionization, and the vacancy sharing among  subshells, together with optimized atomic parameters.
%Individual line and total x-ray production cross sections for L-shell in ultra-thin $Os$ target ionized by 4-6 MeV/u fluorine ions were measured and compared with ECPSSR and ECUSAR theories using single- and multiple-hole fluorescence as well as Coster-Kronig yields. The fluorescence yields are increased approximately by 25-45\%, where as the Coster-Kronig yields are reduced approximately by 10-15\%. Though multiple ionization effects change the theoretical cross section towards the measured ones, but still both the theories are seen to underestimate the measured data to a certain extent, which is not expected at all for  ultra-thin targets. Nevertheless, the measured data get closer to the  ECUSAR theory if L-shell ionization caused by the LK electron capture is taken into account.  K shell vacancies in the fluorine ions are due to the charge state states of 8 and 9+ as estimated by a Fermi gas model. This work demonstrates a case that heavy ion induced innershell ionization on heavy atoms can be understood with the ionization and capture theories in tandem. 
\end{abstract}
\maketitle
%%%%%%%%%%%%%

%%%%%%%%%%%%%%%%%%%%%%%%%%%%%%%%%%%%%%%%%%%%%%%%%%%
\section{Introduction}
\label{S:1}

The measurement of emitted x-rays from targets has resulted in major advances in radiation physics~\cite{satoh2015development}, plasma physics~\cite{sharma2016experimental}, atomic and nuclear physics~\cite{dyson2005x}, and the particle-induced x-ray emission (PIXE) technique~\cite{antoszewska2015modification, gillespie2015advances}. Thus far, the PIXE method has used light ions such as protons or alphas~\cite{johansson1970x, bertol2015proton,garcia1970inner,joseph2013measurement,zhou2013k,miranda2013measurement,miranda2014experimental,mohan2014cross}; however, there is an increasing interest to employ heavy ions since their cross sections are larger and have, thereby, better sensitivity~\cite{siegele1999ansto}. Nevertheless, this potentiality is discouraged by discrepancies observed between the theories and experiments. Although these inconsistencies are often attributed to multiple ionization phenomena~\cite{raju2004multiple,uchai1985x,murillo2016shell}, they do not account for all the discrepancies observed, for example, in experiments with an 8-36 MeV Si-ion beam on targets of Au, Bi, Th, and U with thicknesses between 12 and 40 $\mu g/cm^2$~\cite{fijal2008coupling}. On such occasions, theoretical approaches have been modified to include the $L$-subshell coupling effect as well as the saturation of the binding effect at the united atom limit in addition to the multiple ionization~\cite{fijal2008coupling,pajek2015x}. Even though closeness between the experiments and theory is achieved, differences remain, suggesting that other physical processes are involved. 

It is well-known that for asymmetric collisions, $Z_P/Z_T<1$, the direct ionization (DI) is dominant, whereas for symmetric collisions, $Z_P/Z_T\approx 1$, the electron capture (EC) process becomes increasingly important. Although the present collisional system --F ions ($Z_P=9$) impinging on Os ($Z_T=76$)-- is asymmetric ($Z_P/Z_T$= 0.1184), the measurements cannot be accounted for without considering the EC contribution to the $L$-shell vacancy production~\cite{chatterjee2021exploring}. 
This phenomenon can be understood by considering the ratio of the projectile velocity $V_P$ to the orbital velocity of the $L$-shell electrons $V_L$. In our full-relativistic calculations, the mean velocities (in a.u.) of the osmium $L_i$ sub-shells are $v_i$= 26.6, 37.7, and 32.8 for $i=1,2,3$, respectively. Therefore, 0.33 $\leq V_P/V_L \leq 0.58$, and the collision is asymmetric but in the slow velocity regime. Hence, we included the EC contribution along with the DI, the multiple ionization, and the vacancy sharing. However, contrary to our expectations, the EC does not account for all the discrepancies found between our measurements and the theoretical descriptions. In the final stage of this work, we found that the presently available atomic parameters are not properly described. We modified these values iteratively until a good agreement between the theory and experiment was achieved.

To reveal the $L$-shell ionization mechanism of the collisions of 4-6 MeV/u fluorine ions on an ultra thin osmium target, we provide experimental details in Section \ref{sec:exp}. The theoretical methods employed to describe the ionization are discussed in Section \ref{sec:theo}. Section \ref{sec:params} addresses the effects of the single- and multiple-hole atomic parameters required for the derivation of the subshell-ionization cross sections from the measured x-ray production cross sections. Section \ref{sec:capture} describes the inclusion of ionization by the LK capture processes, and Section ~\ref{sec:discussions} summarizes the major findings.

%%%%%%%%%%%%%%%%%%%%%%%%%%%%%%%%%%%%%%%%%%%%%%%%%%%%%%%%%%%%%%%%%%%%%%%%
%%%%%%%%%%%%%%%%%%
\section{Experimental details and data analysis}
\label{sec:exp}

\begin{figure}[t]
\centering
\includegraphics[width=8cm,height=8cm]{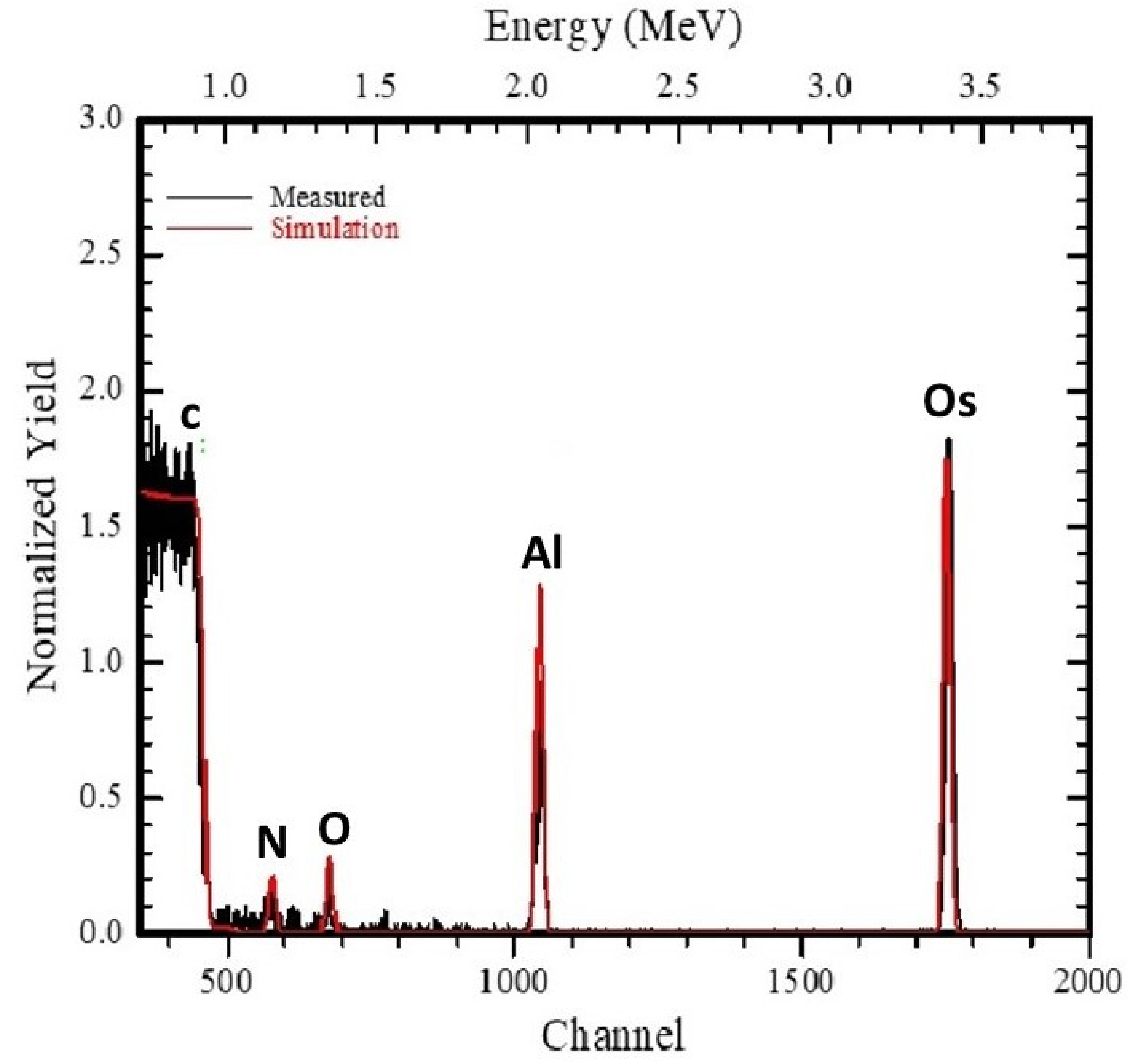}
\caption{The Rutherford  backscattering spectrum of the Os target by $^4$He ion impact.}
\label{RBS} 
\end{figure}

\begin{figure*}
\centering
\includegraphics[width=18cm,height=9cm]{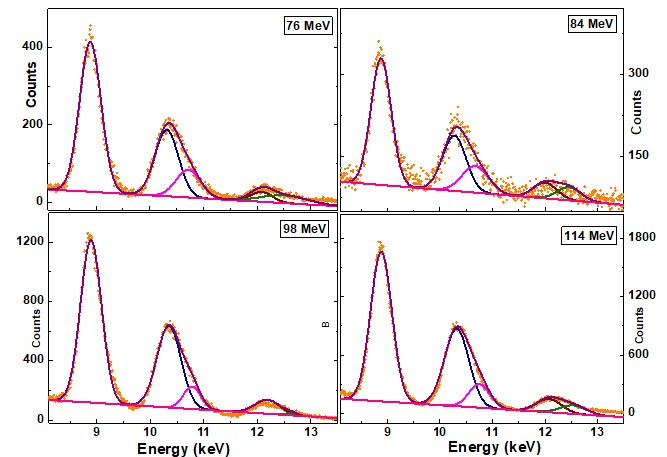}
\caption{The $L$ x-ray spectra of the Os target bombarded by $^{19}$F ions of energies 76, 84, 98, and 114 MeV. The pink 
lines indicate the baseline correction for the background due to Compton scattering.}
\label{SPECTRUM1}
\end{figure*}

\begin{figure}
\centering
\includegraphics[width=7.5cm,height=4cm]{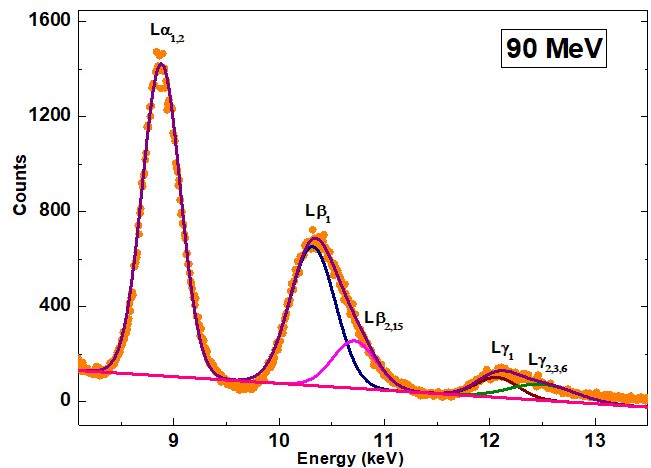}
\caption{Same as Fig.~\ref{SPECTRUM1} but for an impact energy of 90 MeV. Deconvoluted x-ray lines due to different transitions are shown.}
\label{SPECTRUM2}
\end{figure}

The $L$-shell x-ray production cross sections in the Os elements fusing the $^{19}$F ions (charge states $q = 6+$, $7+$, $8+$) in the 76–114 MeV energy range have been measured in the atomic physics beamline at the Inter-University Accelerator Centre, New Delhi. The heavy ions of fluorine --F$^{6+}$ (76 and 84 MeV), F$^{7+}$ (90 MeV) and F$^{8+}$ (98 and 114 MeV)-- were obtained from the 15 UD Pelletron accelerator. The chamber has provision for two silicon surface barrier (SSB) detectors at $\pm$ 7.5$^o$ and two x-ray detectors at 55$^o$ and 125$^o$ to the beam direction, respectively. The target was mounted on a steel ladder forming a 90$^o$ angle to the beam direction. The vacuum inside the chamber was $\sim 10^{-6}$ Torr. The spot of the ion beam at the target had a diameter of approximately 2~mm. The spectra were taken at different positions of each target. Details of the experimental setup and detection system are given by Kumar et al.~\cite{kumar2017shell}. The ultra-thin target of $_{76}$Os was prepared on the polypropylene backing using an ultra-high vacuum deposition setup at IUAC, New Delhi. The thickness of the target was measured using the Rutherford Back-scattering (RBS) method and its spectrum is given in Fig.~\ref{RBS}. The target turned out to be very thin, only 1.09 $\mu$g/cm$^2$. The beam current was kept below 1~nA to avoid pile up effects and damage to the target. The spectra were collected for a sufficiently long time to get good statistical accuracy. The $L$ x-ray spectra of natural Os bombarded by F$^{q+}$ at different projectile energies (76-114 MeV) is shown in Figs.~\ref{SPECTRUM1} and \ref{SPECTRUM2}. The spectra were analyzed with a fitting method considering a Gaussian line shape for the x-ray peaks and a suitable background function. From the figures, it is clear that all major $L$ x-ray components are well resolved by the Si(Li) detector. The details about the data acquisitions and the terms related to the projectile velocity are given by Oswal et al.~\cite{oswal2020experiment}. 

The measured $L$ x-ray production cross sections for the major peaks namely $L_l$, $L_\alpha$, $L_\beta$, and $L\gamma$ were obtained using the following relation 
\begin{eqnarray}
\sigma_x^i= \frac{Y_x^i A sin\theta}{N_A \epsilon n_p t \beta}\,,
\label{eq:1}
\end{eqnarray}
where $Y_x^i$ is the intensity of the $i$th x-ray peak, $A$ is the atomic weight of the target, $\theta$ is the angle between the incident ion beam and the target foil surface, $N_A$ is the Avogadro number, $n_p$ is the number of incident projectiles, $\epsilon$ is the effective efficiency of the x-ray detector, $t$ is the target thickness 1.09 $\mu$g/cm$^2$ and $\beta$ is a correction factor for the absorption of the emitted x-rays inside the target.

The absorption correction factor for the absorption of the emitted $L$ x-rays in the target is written as
\begin{eqnarray}
\beta= \frac{1-\exp^{-\mu t}}{\mu t}\,,
\label{eq:2}
\end{eqnarray}
where $\mu$ is the attenuation coefficient inside the target and its unit is $cm^2/g$~\cite{berger1998xcom}. The value of $\beta$ is $\geq$ 0.99 for the target thickness used in the present measurements. The energy loss calculation using the SRIM code~\cite{ziegler2010srim} for the incident beam within the target suggests negligibly small energy loss for the target thickness and the beam energies used in the present work. The ion beam changes its charge state during its passage through the target.

The role of projectile charge state in this collision regime for 4-6 MeV/u is found to be negligible~\cite{oswal2020experiment}. Integrated charge in a Faraday cup measured by a current integrator has been used to count $N_p$ (see discussion in Section~\ref{sec:discussions}). The energy loss calculation using the SRIM code shows that 76 and 114 MeV fluorine ions lose 2.30 and 1.91 keV in the Os target, respectively. The peak areas $Y_x^i$ are evaluated using the computer program CANDLE~\cite{subramaniam}. This software is an improved version of the Levenburg-Marquardt~\cite{marquardt1963algorithm} non-linear minimization algorithms for the peak fitting. The energy calibration of the detector is performed before and after the in-beam measurements. A semi-empirical fitted relative efficiency curve for the present measurement is available in Oswal et al.~\cite{oswal2020experiment}. 

The percentage error in the measured x-ray production cross sections is about $10-15\%$. This error is attributed to the uncertainties in different parameters used in the analysis, namely the photo peak area evaluation ($\sim\!5\% $ for the $L_\alpha$ x-ray peak and $3\%$ for the other peaks), ion beam current ($\sim\!7\%$), target thickness ($\sim\!3\%$). In the energy region of interest, the error of the absolute efficiency values $\epsilon$ ranges between 5 and 8\%.

%%%%%%%%%%%%%%%%%%%%%%%%%%%%%%%%%%%%%%%%%%%%%%%%%%%%%%%%%%%%%%%%%%%%%%%%
\section{Ionization Theories}
\label{sec:theo}
% Section III

To calculate the direct Coulomb ionization cross section, we have employed (i) the coupled-states relativistic semi-classical approximation (RSCA-CSM), (ii) the shellwise local plasma approximation (SLPA) with fully relativistic electronic structure calculations for Os, and (iii) the ECUSAR theory, which accounts for the energy (E) loss and the Coulomb (C) deflection of the projectile, the perturbed-stationary state effect through a united and separated atom (USA) treatment of the ionization process, and by considering the relativistic (R) nature of the $L$-subshell electronic states. All these models are briefly described below. 

%%%%%%%%%%%%%%%%%%%%%%%%%%%%%%%%%%%%%%%%%%%%%%%%%%%%%%%%%%%%%%%%%%%%%%%%
\subsection{The coupled-states relativistic SCA model (RSCA-CSM)}

In the semi-classical approximation (SCA), the general form of the cross section is expressed as follows
\begin{eqnarray}
\sigma_i=2\pi\int_0^\infty \d b\,
b\sum_f| a_f(t=+\infty)|^2\,,
\label{eq:3}
\end{eqnarray}
where $b$ is the impact parameter, and $a_f(t)$ is the excitation amplitude of the $L_i$ sub-state to a final $f$ state. For the continuum states, the sum means integration over the electron energy. Applying the {\it independent-particle model} approximation, the many-electron excitation amplitudes are replaced by single-electron transition amplitudes. 

The subshell coupling  mechanism is described in a way that a ``mixed'' L state is considered as the initial state, instead of a ``pure'' atomic state~\cite{sarkadi1984higher}. The mixed state evolves in time from the initial $L$-substate as a result of dynamical couplings with the other $L$-substates. The time evolution is governed by the following eight coupled equations 
(in a.u.):
\begin{eqnarray}
\frac{\d a_{n_L^{}}}{\d t}=
-i \sum_{n_L^\prime} {\cal V}_{n_L^{}n_L^\prime}
a_{n_L^\prime}\,,
\label{eq:4}
\end{eqnarray}
with the initial condition
\begin{eqnarray}
a_{n_L^{}}(t=-\infty )=
{\delta}_{n_L^{}i}\,.
\label{eq:5}
\end{eqnarray}
The  excitation from the mixed $L$-substate to the $f$ final state is described by
\begin{eqnarray}
\frac{\d a_f}{\d t}=-i \sum_{n_L^{}}
{\cal V}_{fn_L^{}}a_{n_L^{}}\,.
\label{eq:6}
\end{eqnarray}
%%%%%%%%%%%%%%%%%%%%%%%%%%%%
In Eqs. (\ref{eq:4}) to (\ref{eq:6}), $n_L$ represents the quantum numbers ($l$, $j$, $m_j$) of the L substates. The ${\cal V}_{mk}(t)$ matrix elements for the projectile-target-electron interaction are defined as 
%%%%%%%%%%%%%%%%%%%%%%%%%%%%%%
\begin{eqnarray}
{\cal V}_{mk}(t)=
V_{mk}(t)\exp (i \omega_{mk}t)\,,
\label{eq:7}
\end{eqnarray}
%%%%%%%%%%%%%%%%%%%%%%%%%%%%%%%%%
\begin{eqnarray}
V_{mk}(t)=
\int \d {\bf r}\,\psi_m^{\ast}({\bf r}){-Z_1\over |{\bf r}-{\bf R}(t,b)|}
\psi_k({\bf r})\,,
\label{eq:8}
\end{eqnarray}
%%%%%%%%%%%%%%%%%%%%%%%%%%%%%%%
\begin{eqnarray}
\omega_{mk}=E_m-E_k\,,
\label{eq:9}
\end{eqnarray}
%%%%%%%%%%%%%%%%%%%%%%%%%%%%%%%%%%%%%%%%%
where $Z_1$ is the atomic number of the projectile, ${\bf R}$ is the internuclear vector, and $\psi_j({\bf r})$ and $E_j$
are the one-electron energy eigenstates and eigenvalues of the unperturbed target atom. For ${\bf R(t,b)}$, a Kepler projectile orbit is applied. 

In the present SCA model, the $V_{mk}$ matrix elements are calculated using screened relativistic wave functions for both the bound and continuum states~\cite{lugosi2001calculation}, i.e., it is an RSCA model. Furthermore, since it includes the subshell coupling effects, the full name of the model is coupled-states RSCA, briefly: RSCA-CSM.

We stress that unlike previous works (see, e.g., Refs.~\cite{sarkadi1988subshell,sarkadi1990dynamical}), the integration over the final states in Eq.~(\ref{eq:3}) is complete, and the present RSCA-CSM calculation extends to the entire range of the energy transfer. Additionally, no restrictions are imposed on  the angular momentum of the ionized electron.

%%%%%%%%%%%%%%%%%%%%%%%%%%%%%%%%%%%%%%%%%%%%%%%%%%%%%%%%%%%%%%%%%%%%%%%%
\subsection{ECPSSR/ECUSAR-CSM model}

The ECPSSR model by Brandt and Lapicki~\cite{brandt1981energy} and its evolution into the ECUSAR model~\cite{lapicki2002status} are the most employed theories to describe inner-shell ionization cross sections. They cover an extended energy range, and are the usual input in PIXE codes~\cite{lapicki2002status}. The ECPSSR theory goes beyond  the plane-wave Born approximation (PWBA) by accounting for the energy loss (E), the Coulomb  deflection from a straight-line trajectory and retardation of the projectile (C) and its influence on the unperturbed and non-relativistic atomic orbitals in a perturbed stationary state (PSS) treatment that also accounts for the relativistic (R) nature of the inner shells of heavy target atom. In the ECUSAR theory of~\citet{lapicki2002status}, the PSS treatment of ECPSSR is replaced by the united (U) and separated (S) atom (A) formula (see Eqs. (2) and (3) in~\cite{lapicki2002status}).

Comparing RSCA-CSM to ECPSSR, one has to stress that the former automatically incorporates most of the effects that are included in ECPSSR as corrections. The application of Kepler orbit accounts for the Coulomb-deflection effect. The diagonal matrix elements $V_{mk}$ determine in a first-order approximation the change of the binding energies of the $L$ substates in the presence of the projectile, i.e., they account for the increased binding effect, which is one of the most important consequences of the PSS approach. And, of course, the electronic relativistic effects are exactly taken into account. At the same time, there are differences between the two theories (in addition to the subshell couplings). The ECPSSR model is based on the plane-wave Born approximation (PWBA). Although the PWBA and the straight-line SCA are equivalent, this feature holds only for the hydrogen atom, or hydrogenic ions. For a many-electron atom, it is known that the screening procedure is different for the two theories. It can be shown that the outer screening applied in PWBA leads to an effective potential that approaches the many-electron potential better than the corresponding effective potential in SCA.  Furthermore, the energy loss effect is not included in the RSCA-CSM,  i.e., one does not expect a good performance of the model at very low collision velocities.

In light of the above arguments, the best description of the ion-induced $L$-shell ionization is expected to be given by a modified ECPSSR that includes the subshell coupling effects. Such a model, named ECPSSR-CSM, is obtained by combining ECPSSR and RSCA-CSM in the following way. The original idea behind the consideration of the couplings between the L substates was the vacancy sharing process among the $L$ subshells. The vacancy sharing does not change the total $L$-shell ionization cross section, which is supposed to be described well by ECPSSR. At the same time, the relative subshell ionization cross sections change according to RSCA-CSM. This concept can be expressed by renormalizing the RSCA-CSM subshell cross sections in a way that its sum equals the total ECPSSR cross section (see also~\cite{lapicki2004effects}),
\begin{gather}
\begin{split}
\sigma_{Li} (\mathrm{ECPSSR-CSM}) &= \sigma_{Li} (\mathrm{RSCA -CSM})\\ 
&\quad
\times\frac{\sigma_{\mathrm{Tot}}(\mathrm{ECPSSR-CSM})}{\sigma_{\mathrm{Tot}}(\mathrm{RSCA-CSM})}\,.
\label{eq:10}
\end{split}
\end{gather}
In the present work, we applied the improved version of ECPSSR, the ECUSAR model. According to Eq.~(\ref{eq:10}), the subshell coupling effects are also included in the latter model.

%%%%%%%%%%%%%%%%%%%%%%%%%%%%%%%%%%%%%%%%%%%%%%%%%%%%%%%%%%%%%%%%%%%%%%%%
\subsection{Shellwise local plasma approximation and coupled state model (SLPA-CSM)}

The shellwise local plasma approximation (SLPA)~\cite{montanari2011collective,montanari2013theory} is an ab-initio approach for the calculation of ionization probabilities within the dielectric formalism. Based on the quantum dielectric response theory, which is generally employed to deal with the conduction band of solids, it has been extended to account for the inner-shells by considering the density of target electrons and the binding energies. The SLPA calculates the $j$ subshell ionization cross section of certain target atoms due to the interaction with a projectile (velocity $v_1$ and nuclear charge $Z_1$) as
\begin{gather}
\begin{split}
\sigma_{j}^{\mathrm{SLPA}} =& 2/(\pi v_1^2)\int_{0}^{\infty}\frac{Z_1^2}{p}dp\int_{0}^{p v_1}  d\omega \\ &\qquad\quad
\times\int Im[\frac{-1}{\epsilon(p, \omega, E_j, \delta_j(r)}]\ \vec{dr}\,,
\label{Eq_SLPA}
\end{split}
\end{gather}
with $\epsilon(p,\omega, E_j,\delta_j)$ being the Levine-Louie dielectric function~\cite{levine1982new}, $E_j$ the binding energy, $\delta_j(r)$ the density of the $j$-subshell electrons around the nucleus, and $p(w)$ the momentum (energy) transferred. The $L_i$ subshells of Os were described by performing full relativistic atomic structure calculations. We solved the Dirac equation by implementing the parametric potential method~\cite{Klapisch:67,Klapisch:71,Bar-Shalom:01} and an optimized configuration interaction mixing. The electronic structure computed, which is then used in Eq.~(\ref{Eq_SLPA}), agrees with available experimental values within 1.5\%~\cite{mendez2019relativistic}. It is worth noting that the SLPA has been successfully employed previously by the authors to obtain $L$-shell ionization  cross sections of relativistic targets such as Pt, Ta, W, Pb, Au, Bi, Th and U~\cite{oswal2018x,oswal2020experiment}. %, K-shell ionization    \cite{kadhane2003k,chatterjee2021investigation}, or mean energy loss \cite{montanari2017low,montanari2020stopping}.
%%%%%%%%%%%
%\rrt{[Please see above how ECUSAR-CSM is obtained from ECUSAR and PWBA (first order value of ECUSAR). If you (CCM) can provide us the first order SLPA values of L-subshell cross sections then we can evaluate SLPA-CSM values also.]}

%%%%%%%%%%%%%%%%%%%%%%%%%%%%%%%%%%%%%%%%%%%%%%%%%%%%%%%%%%%%%%%%%%%%%%%%
\subsection{Method for obtaining theoretical x-ray production cross section from theoretical $L$-shell ionization cross section}

The theoretical $L$ x-ray production cross sections for the most commonly resolved $L_l$, $L_{\alpha}$, $L_{\beta}$, $L_{\gamma}$ x-rays are related to the $L_i$ subshell ionization cross sections, $\sigma_{Li}$, as given below
\begin{gather}
\sigma^{x}_{L_l}=[\sigma_{L_1}(f_{13}+(f_{12}f_{23})+\sigma_{L_2}f_{23}+\sigma_{L_3}]\omega_3F_{3l} \,,
\label{eq:13}
\end{gather}
%%%%%
\begin{gather}
\sigma^{x}_{L_{\alpha}}=[\sigma_{L_1}(f_{13}+(f_{12}f_{23})+\sigma_{L_2}f_{23}+\sigma_{L_3}]\omega_3F_{3\alpha} \,,
\label{eq:14}
\end{gather}
%%%%%%%%%%%%%
\begin{gather}
\begin{split}
\sigma^{x}_{L_{\beta}}=\sigma_{L_1}[\omega_1F_{1\beta}+f_{12}\omega_2F_{2\beta}+(f_{13}+(f_{12}f_{23})\omega_3F_{3\beta}]+\\
\sigma_{L_2}(\omega_2F_{2\beta}+f_{23}\omega_3F_{3\beta})+\sigma_{L_3}\omega_3F_{3\beta} \,,
\label{eq:15}
\end{split}
\end{gather}
%%%%%%%%%%%
\begin{gather}
\sigma^x_{L_{\gamma}}=\sigma_{L_1}\omega_1F_{1\gamma}+(\sigma_{L_1}f_{12}+\sigma_{L_2})\omega_2F_{2\gamma} \,,
\label{eq:16}
\end{gather}
where $\sigma_{L_p}^x$ is the x-ray production cross sections of the different $L$ x-ray components, $\sigma_{L_i}$ is the ionization cross sections for the $L_i$ subshell, $\omega_i$ is the fluorescence yields of the $L_i$ subshells, $f_{ij} (i < j)$ is the CK yields for the CK transition between the $L_i$ and $L_j$ subshells, and $F_{ip}$ is the fractional radiative emission rates, with $i= 1,\,2,\,3$ and $p =l,\,\alpha,\,\beta,\,\gamma$. The theoretical $L$ x-ray production cross sections were calculated by combining the $L_i$ ionization cross sections obtained by the different $L$-shell ionization models in Eqs. (\ref{eq:13}) to (\ref{eq:16}).

%%%%%%%%%%%%%%%%%%%%%%%%%%%%%%%%%%%%%%%%%%%%%%%%%%%
\begin{table*}[t]
\caption{The fluorescence and CK yields for the singly-ionized Os denoted by superscript 0, and for the multiply-ionized
Os without superscript. The values listed for the singly-ionized atom were taken from the compilation of Campbell~\cite{campbell2003fluorescence}. The values denoted by superscript $m$ are optimized parameters that resulted in the best agreement between the theoretical and experimental $L$-subshell ionization cross sections (see text).} 
\label{CK_W}
\begin{tabular}{|l|l|l|l|l|l|l|l|l|l|l|l|l|l|}
\hline
\multicolumn{2}{|l|}{\begin{tabular}[c]{@{}l@{}}Atomic Number\\           \hspace{0.9cm}(Z)\end{tabular}} & \multicolumn{6}{l|}{\hspace{1.5cm}Fluorescence Yield}                                                          & \multicolumn{6}{l|}{\hspace{2.1cm}CK Yield}                                                                                                                          \\ \hline
\multicolumn{14}{|l|}{\hspace{6.6cm}SI}                                                                                                                                                                                                                                                                                                                               \\ \hline
\multicolumn{2}{|l|}{\multirow{2}{*}{76}}                                                   & $\omega_1^0$ & $\omega_1^{0m}$ & $\omega_2^0$ & $\omega_2^{0m}$ & $\omega_3^0$ & $\omega_3^{0m}$ & $f_{12}^0$ & $f_{12}^{0m}$ & $f_{13}^0$ & $f_{13}^{0m}$ & $f_{23}^0$ & $f_{23}^{0m}$ \\ \cline{3-14} 
\multicolumn{2}{|l|}{}                                                                      & 0.15         & 0.279           & 0.318        & 0.34            & 0.282        & 0.305           & 0.07       & 0.08                                & 0.33       & 0.36                                & 0.13       & 0.17                                \\ \hline
\multicolumn{13}{|l|}{\hspace{6.6cm}MI}                                                                                                                                                                                                                                                                                         &                                     \\ \hline
E (MeV)                                       & $Q_m$                                       & $\omega_1$   & $\omega_1^m$    & $\omega_2$   & $\omega_2^m$    & $\omega_3$   & $\omega_3^m$     & $f_{12}$   & $f_{12}^m$                          & $f_{13}$   & $f_{13}^m$                          & $f_{23}$   & $f_{23}^m$                          \\ \hline
76                                            & 8.73                                        & 0.196        & 0.348           & 0.392        & 0.416           & 0.352        & 0.377           & 0.0367     & 0.0441                              & 0.173      & 0.19                                & 0.0682     & 0.0892                              \\ \hline
84                                            & 8.74                                        & 0.190        & 0.34            & 0.383        & 0.407           & 0.344        & 0.369           & 0.0394     & 0.0473                              & 0.186      & 0.204                               & 0.0732     & 0.0957                              \\ \hline
90                                            & 8.75                                        & 0.187        & 0.336           & 0.378        & 0.402           & 0.339        & 0.364           & 0.0411     & 0.0493                              & 0.194      & 0.213                               & 0.0763     & 0.0998                              \\ \hline
98                                            & 8.76                                        & 0.184        & 0.33            & 0.373        & 0.396           & 0.333        & 0.359           & 0.0431     & 0.0518                              & 0.203      & 0.224                               & 0.0801     & 0.105                               \\ \hline
114                                           & 8.78                                        & 0.178        & 0.322           & 0.364        & 0.388           & 0.325        & 0.350           & 0.0464     & 0.0570                              & 0.219      & 0.241                               & 0.0862     & 0.113                               \\ \hline
\end{tabular}
\end{table*}
%%%%%%%%%%%%%%%%%%%%%%%%%%%%%%%%%%%%%%%%%%%%%%%%
%%%%%%%%%%%%%%%%%%%%%%%%%%%%%%%%%%%%%%%%%%%%%%%%%%%%%%%%%%%%%%%%%%%%%%%%
\section{EFFECT OF THE MULTIPLE VACANCIES ON THE ATOMIC PARAMETERS USED FOR THE CONVERSION OF THE IONIZATION CROSS SECTIONS TO X-RAY PRODUCTION CROSS SECTIONS}
\label{sec:params}

Multiple vacancies in the target atom change the atomic parameters: the fluorescence yields and the CK yields, which in turn alter the x-ray production cross sections. In the present work, single-hole fluorescence $\omega_i^0$ and CK yields $f_{ij}^0$~\cite{campbell2003fluorescence}, were corrected for multiple ionization using a model prescribed by Lapicki et al.~\cite{lapicki1986multiple}. Each electron in a manifold of the outer subshells is ionized with a probability $P$, which is calculated using Eq.~(A3) from~\cite{lapicki1986multiple}, and replacing the projectile atomic number $Z_P$ by its charge state $q$~\cite{mehta1993shell},
\begin{eqnarray}
P=\frac{q^2}{2\beta V_p^2}(1-\frac{\beta}{4V_p^2})\,,
\label{eq:17}
\end{eqnarray}
with $\beta= 0.9$. For the charge state $q$, we take the incident charge state of the projectile. The $\omega_i^0$ values corrected for the simultaneous multiple ionization (SMI) in the outer subshells are given by
\begin{eqnarray}
\omega_i=\omega_i^0[1-P(1-\omega_i^0)]^{-1}\,,
\label{eq:18}
\end{eqnarray}
while the $f_{ij}$ values for multiple ionization are given by
\begin{eqnarray}
f_{ij}=f_{ij}^0[1-P]^{2}\,.
\label{eq:19}
\end{eqnarray}

Note that the fractional rates $F_{ip}$ remain unchanged because both the partial and the total non-radiative widths are narrowed by identical factors. According to Eqs. (\ref{eq:18}) and (\ref{eq:19}), the single-hole fluorescence and CK yields depend on the energy and charge state of the projectile ion. The fluorescence and the CK yields for singly- and multiply-ionized  Os is given in Table \ref{CK_W}. It is clear from this table that in the extreme the $L_i$ subshell fluorescence yields are enhanced by $\sim$30\%, and the CK yields are reduced up to $\sim$50\% from the single-hole to the multiple-hole atom in Os. Note that the use of different sets of atomic parameters can change the x-ray production cross section by $\sim$30\% or more. Recent values of $\omega_i^0$, and $f_{ij}^0$, compiled by Campbell~\cite{campbell2003fluorescence} for the elements with $25\leq Z \leq 96$, have been used in the present work for singly-ionized atoms.
%%%%%%%%%%%%%
%\indent L-shell line and total x-ray production cross sections, at corresponding energies and incident charge state of the fluorine ions, are listed in Table \ref{CK_W}.  Although the connection between observed L x-ray lines and calculated $L_i$ subshell ionization cross sections depends on a combination of intra-shell coupling and inner shell multiple ionization effects    \cite{sarkadi1986l3,sarkar1995importance}, the data for ionization of comparably heavy targets as ours but by significantly slower 4-8 MeV carbon ions    \cite{lapicki2004effects} show that multiple-ionization is more effective. While both effects subside in the 4-6 MeV/u range of the present experiment, the effect of the intra-shell coupling is overshadowed by multiple ionization    \cite{lapicki2004effects}. Thus in Table 1, ignoring the negligible effect the intra-shell coupling, all measured cross sections are compared to the predictions of the ECPSSR    \cite{brandt1981energy} and ionization theories converted to the x-ray production cross sections using multiple-hole atomic parameters calculated with Eqs.\ref{eq:5}-\ref{eq:7}.
%%%%%%%%%%%%%%%%%%%%%%%%%%%%%%%%%%%%%%%%%%%%%%%%%%%%%%%%%%%%%%%%%%%%%%%%

%%%%%%%%%%%%%%%%%%%%%%%%%%%%%%%%%%%%%%%%%%%%%%%%%%%%%%%%%%%%%%%%%%%%%%%%
\section{EFFECT OF THE $L$-$K$ ELECTRON CAPTURE ON THE $L$-SHELL IONIZATION CROSS SECTION}
\label{sec:capture}

To calculate the $L$-$K$ electron capture cross sections, we used the theory of Lapicki and Losonsky~\cite{lapicki1977electron}, which is based on the Oppenheimer-Brinkman-Kramers (OBK) approximation~\cite{R} with binding and Coulomb deflection corrections at low velocities. Neglecting the change in the binding energy of the $K$ shell electron of the projectile with one versus two $K$ shell vacancies, a statistical scaling is used to calculate the electron transfer cross section for the case of one projectile $K$-shell vacancy, $\sigma_{L \rightarrow K}$, resulting in $\sigma_{L \rightarrow 2K}/2$, where $\sigma_{L \rightarrow 2K}$ is the production cross section for two projectile $K$-shell vacancies. In the present experimental condition, $v_1$ ranges between 12.39 and 15.17, while $v_{2L}= Z_{2L}/n_2=35.925$ (in a.u.), $n_2$ and $n_1$ are the principal quantum numbers of $L$ and $K$ shell electrons of the target and the projectile atom, respectively. Following~\citet{lapicki1977electron}, $\sigma_{L \rightarrow 2K}$ can be obtained as
\begin{equation}
\sigma_{L \rightarrow 2k}=\frac{1}{3}\sigma_{L\rightarrow 2k}^{\mathrm{OBK}}(\theta_L),\theta_k=\frac{E_L}{v_{2L}^2\times13.6}\,,
\label{KKCap} 
\end{equation}
with $Z_{2L}=Z_T-4.15$ and
\begin{equation}
 \sigma_{L\rightarrow 2k}^{\mathrm{OBK}}(\theta_k)= \frac{2^9}{5v_1^2}\pi a_0^2 n_1^2 \frac{(v_{1k}v_{2L})^5 Z_P 10^{24}}{[v_{1k}^2+(v_1^2+v_{2L}^2-v_{1k}^2)^2/4v_1^2]^5}\,,
\end{equation}
where $E_L$ is the binding energy of $L$-shell electron of the target (in eV), and the parameters $a_0$, $v_{1k}$, $Z_p$, and $Z_T$ are the Bohr radius, the $K$-shell orbital velocity of the projectile ion, atomic number of the projectile and the target atom, respectively.

%%%%%%%%%%%%%%%%%%%%%%%%%%%%%%%%%%%%%%%%%%%%%%%%%%%%%%%%%%%%%%%%%%%%%%%%

\begin{figure}
\centering
\includegraphics[width=7.5cm,height=5cm]{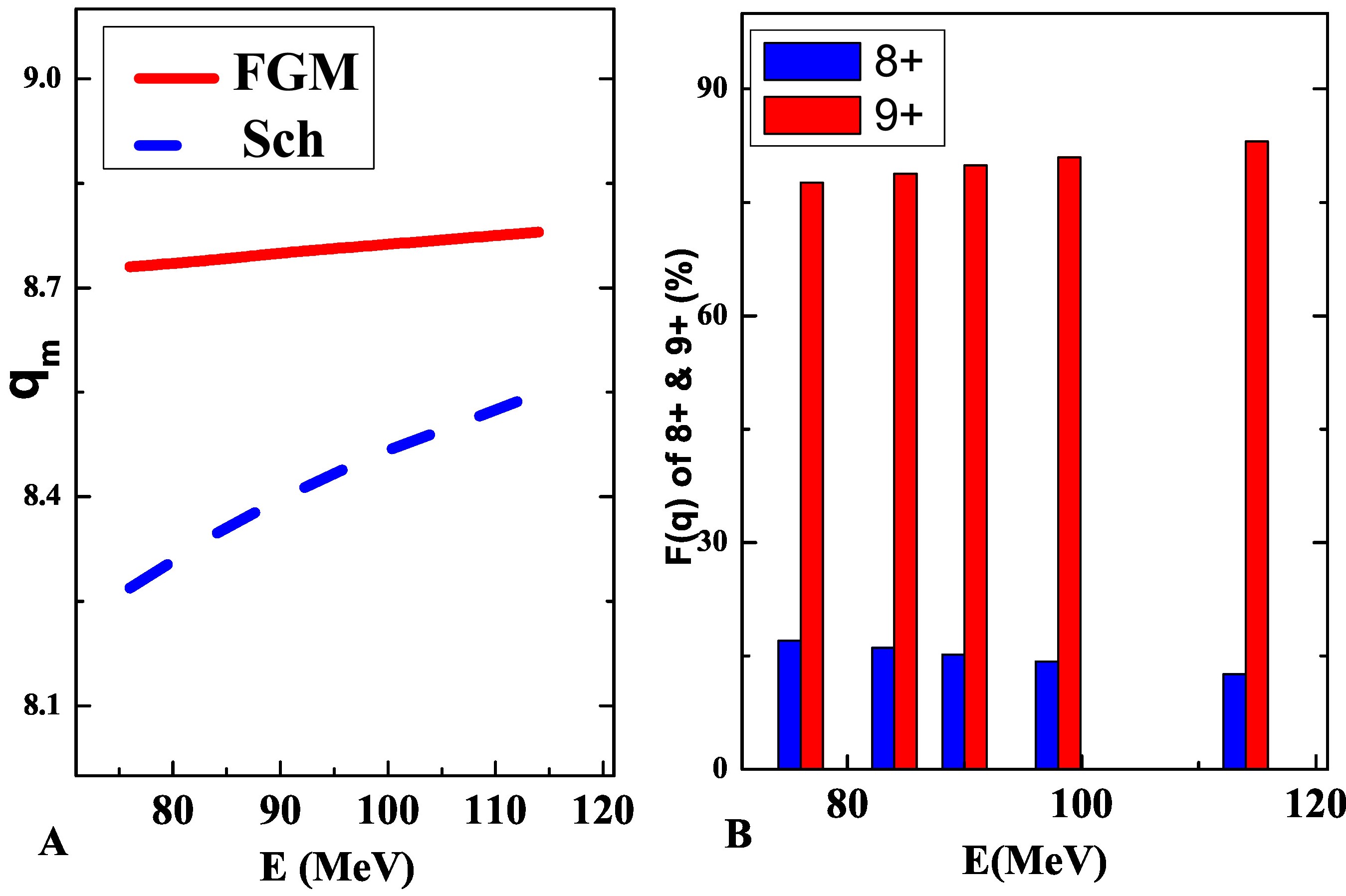}
\caption{A. The mean charge state of the $^{19}$F ion inside the Os target as predicted by the Fermi Gas Model (FGM)~\cite{brandt1973dynamic} and the same outside the target as predicted by the Schiwietz model (Sch)~\cite{schiwietz2001improved} versus the incident energies. B. The charge-state fraction $F(q)$ chart for $q=8+$ and $9+$ inside the target as a function of the beam energy.}
\label{Q-BAR-CSD} 
\end{figure}

%%%%%%%%%%%%%%%%%%%%%%%%%%%%%%%%%%%%%%%%%%%%%%%%%%%%%%%%%%%%%%%%%%%%%%%%

\begin{figure*}
\includegraphics[width=7.5cm,height=6cm]{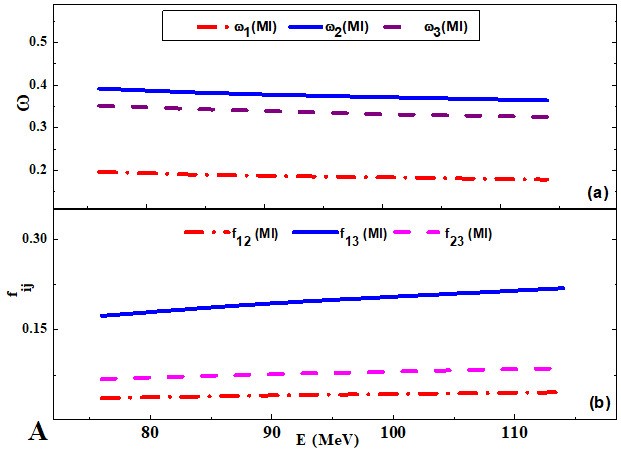}
\includegraphics[width=7.5cm,height=6cm]{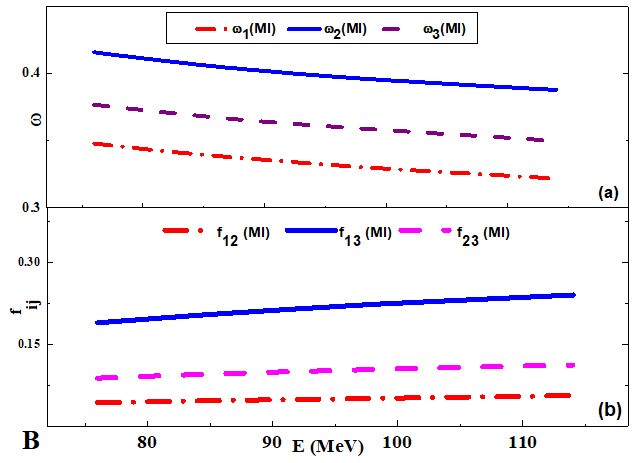}
\caption{A. The effect of the simultaneous multiple ionization on (a) the fluorescence and (b) the Coster-Kronig yields. The plotted data are results of calculations using Eqs. (16) to (18), where the atomic parameters for the single ionization condition were taken from~\cite{campbell2003fluorescence}. B. The same as A, but using such atomic parameters for the single ionization condition which were optimized in order to have a good agreement between the experimental and theoretical $L_i$-subshell ionization cross sections.}
\label{CK-W-GRAPH}
\end{figure*}

Due to certain charge state distribution inside the target, the effective capture contribution will be $F(q)\times \sigma_{L\rightarrow 2k}^{\mathrm{OBK}}(\theta_L)$, for $q=8+$ and $9+$. Here, $F(q)$ is the charge state fraction for a specific $q$. To obtain $F(q)$, we employed a two-fold procedure. First, we used a Fermi-gas-model based empirical formula for the determination of the mean charge state, $q_m$, inside the target~\cite{brandt1973dynamic}  
\begin{equation}
q_m = Z_P(1- {v_F\over v_1})\,,
\end{equation}
where $z_1$ and $v_F$ are the projectile atomic number and the Fermi velocity of target electrons, respectively. The value of $v_F$ for Os is $8.19\times10^6$ cm/sec. To showcase the difference between the ionization of the projectile ion inside and outside the target, we displayed the $q_m$ as predicted by the Fermi-gas-model~\cite{brandt1973dynamic} and by the Schiwietz model~\cite{schiwietz2001improved} in Fig.~\ref{Q-BAR-CSD}(A). This contrasting picture is governed by the solid surface~\cite{sharma2019disentangling,nandi2008formation}. The charge state of the heavy projectiles inside the solid target is higher than for those outside. This feature has been described in detail by~\citet{chatterjee2021exploring}. In the second part of the procedure, the $q_m$ values inside the target are substituted by a Lorentzian charge state distribution~\cite{sharma2019disentangling} to obtain the $F(q)$ as follows
\begin{equation}
F(q)=\frac{1}{2\pi}\frac{\Gamma}{(q-q_m)^2+(\Gamma/2)^2} \,\,\,\, \:\text{and} \,\,\,\, \:\sum_q F(q)=1 \,,
\label{CSF}
\end{equation}
where the distribution width $\Gamma$ is taken from Novikov and Teplova~\cite{novikov2014methods} as follows
\begin{equation}
\centering
\Gamma(x)= C\left[1-\exp(-x^\alpha)\right]\left\{1-\exp\left[-(1-x)^\beta\right]\right\}\label{Gamma}\,,
\end{equation}
being $x=q_m/z,\,\, \alpha=0.23,\,\, \beta=0.32$ and \mbox{$C=2.669-0.0098\,Z_2+0.058\,Z_1+0.00048\,Z_1\,Z_2$}. The $F(q)$ for $q=8+$ and $9+$ are displayed with a bar chart in Fig.~\ref{Q-BAR-CSD}(B).

%%%%%%%%%%%%%%%%%%%%%%%%%%%%%%%%%%%%%%%%%%%%%%%%%%%%%%%%%%%%%%%%%%%%%%%%
%%%%%%%%%%%%%%%%%%
\begin{table}
\centering
\caption{The $L_1, L_2, L_3,$ and $L_{\mathrm{Tot}}$ ionization cross sections (kilo-barn) for $^{19}$F on Os as a function of energy (MeV). TH1, TH2, and TH3 denote the ECUSAR-CSM, the RSCA-CSM, and the SLPA model, respectively, while EC denotes the electron capture.} 
\label{CROSS-SECTION_L123}
\begin{tabular}{|l|l|l|l|l|l|l|l|}
\hline
\begin{tabular}[c]{@{}l@{}}\hspace{0.1cm}E\end{tabular}  & \begin{tabular}[c]{@{}l@{}}\hspace{0.1cm}Expt. \\    \hspace{0.09cm} ($\sigma_{Li}$)\end{tabular} &  \begin{tabular}[c]{@{}l@{}}\hspace{0.1cm}TH1\\        \end{tabular} & \begin{tabular}[c]{@{}l@{}}\hspace{0.1cm}TH1\\             \hspace{0.1cm} +\\     \hspace{0.1cm}EC\\  \end{tabular} & \begin{tabular}[c]{@{}l@{}}\hspace{0.1cm}TH2\\ \end{tabular} & \begin{tabular}[c]{@{}l@{}}\hspace{0.1cm}TH2\\ \hspace{0.1cm}+\\     \hspace{0.1cm}EC\\ \end{tabular} &
\begin{tabular}[c]{@{}l@{}}\hspace{0.1cm}TH3\\  \end{tabular}
&
\begin{tabular}[c]{@{}l@{}}\hspace{0.1cm}TH3\\ \hspace{0.1cm}+\\     \hspace{0.1cm}EC\\ \end{tabular}\\ \hline
\multicolumn{8}{|l|}{\hspace{3.4cm}$i=1$}                                                                                                 \\ \hline
\hspace{0.1cm}76      
           & \hspace{0.1cm}4.7                                                                             
           & \hspace{0.1cm}4.0                                                                
           & \hspace{0.1cm}4.3                                                                               & \hspace{0.1cm}1.9 
           & \hspace{0.1cm}2.2    & \hspace{0.1cm}2.7  & \hspace{0.1cm}3.0                                                                       \\ \hline
\hspace{0.1cm}84      
           & \hspace{0.1cm}6.6              
           
          &   \hspace{0.1cm}5.5                                                                               & \hspace{0.1cm}5.9                                                                               & \hspace{0.1cm}2.7  
           &  \hspace{0.1cm}3.1   &  \hspace{0.1cm}3.7 & \hspace{0.1cm}4.1                                                                         \\ \hline
\hspace{0.1cm}90      
            & \hspace{0.1cm}8.4                           
            & \hspace{0.1cm}6.7                                                                               & \hspace{0.1cm}7.2                                                                               & \hspace{0.1cm}3.4 
            & \hspace{0.1cm}3.9   & \hspace{0.1cm}4.5  & \hspace{0.1cm}5.0                                                                       \\ \hline
\hspace{0.1cm}98      
            & \hspace{0.1cm}9.1                                                    
            & \hspace{0.1cm}8.4                                                                               & \hspace{0.1cm}9.1                                                                               & \hspace{0.1cm}4.3 
            & \hspace{0.1cm}5.0   & \hspace{0.1cm}5.6  & \hspace{0.1cm}6.3                                                                       \\ \hline
\hspace{0.1cm}114     
            & \hspace{0.1cm}13.9                                             
            & \hspace{0.1cm}11.8                                                                              & \hspace{0.1cm}12.8  
            & \hspace{0.1cm}6.3   
            & \hspace{0.1cm}7.2   & \hspace{0.1cm}8.3  & \hspace{0.1cm}9.3  
    \\ \hline
\multicolumn{8}{|l|}{\hspace{3.4cm}$i=2$}                                                                                                 \\ \hline
\hspace{0.1cm}76      
            & \hspace{0.1cm}6.1 
            & \hspace{0.1cm}5.0                                                                               & \hspace{0.1cm}5.3                                                                               & \hspace{0.1cm}2.4 
            & \hspace{0.1cm}2.7  & \hspace{0.1cm}3.9  & \hspace{0.1cm}4.2                                                                        \\ \hline
\hspace{0.1cm}84      
            & \hspace{0.1cm}7.1                                       
            & \hspace{0.1cm}6.2                                                                               & \hspace{0.1cm}6.6                                                                               & \hspace{0.1cm}3.1                   
            &  \hspace{0.1cm}3.5   & \hspace{0.1cm}4.9  & \hspace{0.1cm}5.3                                                                       \\ \hline
\hspace{0.1cm}90      
            & \hspace{0.1cm}7.9       
            & \hspace{0.1cm}7.2                                                                               & \hspace{0.1cm}7.7                                                                               & \hspace{0.1cm}3.6               
            & \hspace{0.1cm}4.1  & \hspace{0.1cm}5.7  & \hspace{0.1cm}6.2                                                                        \\ \hline
\hspace{0.1cm}98      
            & \hspace{0.1cm}9.5                                                                       
            & \hspace{0.1cm}8.5                                                                               & \hspace{0.1cm}9.1                                                                               & \hspace{0.1cm}4.4                                                
            & \hspace{0.1cm}5.0  & \hspace{0.1cm}6.9  & \hspace{0.1cm}7.5                                                                       \\ \hline
\hspace{0.1cm}114     
            & \hspace{0.1cm}14.4                                                                                   
            & \hspace{0.1cm}11.2                                                                              & \hspace{0.1cm}12.2                                                                              & \hspace{0.1cm}6.0                                               
            & \hspace{0.1cm}7.0  & \hspace{0.1cm}9.4  & \hspace{0.1cm}10.4                                                                   \\ \hline
\multicolumn{8}{|l|}{\hspace{3.4cm}$i=3$}                                                                                                 \\ \hline
\hspace{0.1cm}76     
            & \hspace{0.1cm}20.3                                                                                  
            & \hspace{0.1cm}17.3                                                                              & \hspace{0.1cm}17.8                                                                              & \hspace{0.1cm}8.3                                                 
            & \hspace{0.1cm}8.8  & \hspace{0.1cm}14.7  & \hspace{0.1cm}15.2                                                                      \\ \hline
\hspace{0.1cm}84      
            & \hspace{0.1cm}26.1                                                                                    
            & \hspace{0.1cm}21.0                                                                              & \hspace{0.1cm}21.7                                                                              & \hspace{0.1cm}10.3                                                  
            & \hspace{0.1cm}11.0  & \hspace{0.1cm}17.7  & \hspace{0.1cm}18.4                                                                      \\ \hline
\hspace{0.1cm}90      
            & \hspace{0.1cm}27.3                                                                                
            & \hspace{0.1cm}23.8                                                                              & \hspace{0.1cm}24.6                                                                              & \hspace{0.1cm}12.0                                                   
            & \hspace{0.1cm}12.7  & \hspace{0.1cm}20.1  & \hspace{0.1cm}20.9                                                         
    \\ \hline
\hspace{0.1cm}98      
            & \hspace{0.1cm}30.4                                                                                       
            & \hspace{0.1cm}27.7                                                                              & \hspace{0.1cm}28.9                                                                              & \hspace{0.1cm}14.2                                                 
            & \hspace{0.1cm}15.4  & \hspace{0.1cm}23.4  & \hspace{0.1cm}24.6                                                                      \\ \hline
\hspace{0.1cm}114     
            & \hspace{0.1cm}39.2                                                                                   
            & \hspace{0.1cm}35.4                                                                              & \hspace{0.1cm}37.1                                                                              & \hspace{0.1cm}18.8                                                     
            & \hspace{0.1cm}20.5  & \hspace{0.1cm}30.3  & \hspace{0.1cm}32.0                                                                       \\ \hline
\multicolumn{8}{|l|}{\hspace{3.4cm}$L_{\mathrm{Tot}}$}                                                            \\ \hline
\hspace{0.1cm}76      
            & \hspace{0.1cm}31.1                                                                                             
            & \hspace{0.1cm}26.3                                                                              & \hspace{0.1cm}27.4                                                                              & \hspace{0.1cm}12.7                                                  
            & \hspace{0.1cm}13.8  & \hspace{0.1cm}21.3 & \hspace{0.1cm}22.4                                                                       \\ \hline
\hspace{0.1cm}84      
            & \hspace{0.1cm}39.8                                                                                   
            & \hspace{0.1cm}32.7                                                                              & \hspace{0.1cm}34.2                                                                              & \hspace{0.1cm}16.1                                                   
            & \hspace{0.1cm}17.6  & \hspace{0.1cm}26.3  & \hspace{0.1cm}27.8                                                                        \\ \hline
\hspace{0.1cm}90      
            & \hspace{0.1cm}43.6                                                                                   
            & \hspace{0.1cm}37.6                                                                              & \hspace{0.1cm}39.4                                                                              & \hspace{0.1cm}19.0                                                
            & \hspace{0.1cm}20.7  & \hspace{0.1cm}30.3  & \hspace{0.1cm}32.1                                                                       \\ \hline
\hspace{0.1cm}98      
            & \hspace{0.1cm}49.0                                  
            & \hspace{0.1cm}44.5                                                                              & \hspace{0.1cm}47.0                                                                              & \hspace{0.1cm}22.9                                                
            & \hspace{0.1cm}25.4  & \hspace{0.1cm}35.9  & \hspace{0.1cm}38.4                                                                       \\ \hline
\hspace{0.1cm}114     
            & \hspace{0.1cm}67.5                                                                               
            & \hspace{0.1cm}58.4                                                                              & \hspace{0.1cm}61.1                                                                              & \hspace{0.1cm}31.0                                      
            & \hspace{0.1cm}34.7  & \hspace{0.1cm}48.0  & \hspace{0.1cm}51.7                                                                       \\ \hline
\end{tabular}
\end{table}

%NEW TABLE_PRODUCTION CROSS-SECTION
%%%%%%%%%%%%%%%%%%%%%%%%%%%%%%%%%%%%%%%%%%%%%%%%%%%%%
\begin{table}
\centering
\caption{ The $L_\alpha, L_\beta, L_\gamma,$ and $L_{Tot}$ x-ray production cross sections for $^{19}$F on Os as a function of energy. Same notation and units as Table.~\ref{CROSS-SECTION_L123}.}
\label{CROSS-SECTION}
\begin{tabular}{|l|l|l|l|l|l|l|l|}
\hline
\begin{tabular}[c]{@{}l@{}}\hspace{0.1cm}E \\ \end{tabular} & \begin{tabular}[c]{@{}l@{}}\hspace{0.1cm}Expt. \\    ($\sigma^x_L$) \end{tabular} & \begin{tabular}[c]{@{}l@{}}\hspace{0.1cm}TH1\\ \end{tabular} & \begin{tabular}[c]{@{}l@{}}\hspace{0.1cm}TH1\\             \hspace{0.1cm} +\\     \hspace{0.1cm}EC\\  \end{tabular} & \begin{tabular}[c]{@{}l@{}}\hspace{0.1cm}TH2\\ \end{tabular} & \begin{tabular}[c]{@{}l@{}}\hspace{0.1cm}TH2\\ \hspace{0.1cm}+\\     \hspace{0.1cm}EC\\ \end{tabular} &
\begin{tabular}[c]{@{}l@{}}\hspace{0.1cm}TH3\\  \end{tabular}
&
\begin{tabular}[c]{@{}l@{}}\hspace{0.1cm}TH3\\ \hspace{0.1cm}+\\     \hspace{0.1cm}EC\\ \end{tabular}\\ \hline
\multicolumn{8}{|l|}{\hspace{3.6cm}Ll}                                                                                                                                                                                                                                                                                                                                                                                                                \\ \hline
\hspace{0.1cm}76      & \hspace{0.1cm}0.27                                                                    & \hspace{0.1cm}0.27                                                        & \hspace{0.1cm}0.28                                                                                               & \hspace{0.1cm}0.13                                                     & \hspace{0.1cm}0.14                                                                                          & \hspace{0.1cm}0.22  & \hspace{0.1cm}0.23   \\ \hline
\hspace{0.1cm}84      & \hspace{0.1cm}0.37                                                                    & \hspace{0.1cm}0.32                                                        & \hspace{0.1cm}0.33                                                                                               & \hspace{0.1cm}0.16                                                     & \hspace{0.1cm}0.17                                                                                          & \hspace{0.1cm}0.27  & \hspace{0.1cm}0.28   \\ \hline
\hspace{0.1cm}90      & \hspace{0.1cm}0.44                                                                    & \hspace{0.1cm}0.36                                                        & \hspace{0.1cm}0.37                                                                                               & \hspace{0.1cm}0.18                                                     & \hspace{0.1cm}0.19                                                                                          & \hspace{0.1cm}0.30    & \hspace{0.1cm}0.31   \\ \hline
\hspace{0.1cm}98      & \hspace{0.1cm}0.45                                                                     & \hspace{0.1cm}0.42                                                        & \hspace{0.1cm}0.44                                                                                               & \hspace{0.1cm}0.21                                                     & \hspace{0.1cm}0.23                                                                                          & \hspace{0.1cm}0.35  & \hspace{0.1cm}0.37   \\ \hline
\hspace{0.1cm}114     & \hspace{0.1cm}0.63                                                                     & \hspace{0.1cm}0.53                                                        & \hspace{0.1cm}0.56                                                                                               & \hspace{0.1cm}0.28                                                     & \hspace{0.1cm}0.31                                                                                          & \hspace{0.1cm}0.45  & \hspace{0.1cm}0.47   \\ \hline
\multicolumn{8}{|l|}{\hspace{3.6cm}$L_\alpha$}                                                                                                                                                                                                                                                                                                                                                                                                                \\ \hline
\hspace{0.1cm}76      & \hspace{0.1cm}5.53                                                                     & \hspace{0.1cm}5.52                                                         & \hspace{0.1cm}5.70                                                                                                 & \hspace{0.1cm}2.65                                                      & \hspace{0.1cm}2.82                                                                                           & \hspace{0.1cm}4.64   & \hspace{0.1cm}4.82    \\ \hline
\hspace{0.1cm}84      & \hspace{0.1cm}7.67                                                                     & \hspace{0.1cm}6.63                                                         & \hspace{0.1cm}6.87                                                                                                & \hspace{0.1cm}3.25                                                      & \hspace{0.1cm}3.49                                                                                           & \hspace{0.1cm}5.52   & \hspace{0.1cm}5.76    \\ \hline
\hspace{0.1cm}90      & \hspace{0.1cm}9.32                                                                     & \hspace{0.1cm}7.47                                                         & \hspace{0.1cm}7.75                                                                                                & \hspace{0.1cm}3.77                                                      & \hspace{0.1cm}4.02                                                                                           & \hspace{0.1cm}6.23   & \hspace{0.1cm}6.50    \\ \hline
\hspace{0.1cm}98      & \hspace{0.1cm}9.38                                                                     & \hspace{0.1cm}8.64                                                          & \hspace{0.1cm}9.05                                                                                                & \hspace{0.1cm}4.43                                                      & \hspace{0.1cm}4.84                                                                                           & \hspace{0.1cm}7.2    & \hspace{0.1cm}7.60    \\ \hline
\hspace{0.1cm}114     & \hspace{0.1cm}13.00                                                                     & \hspace{0.1cm}10.90                                                         & \hspace{0.1cm}11.5                                                                                                & \hspace{0.1cm}5.82                                                      & \hspace{0.1cm}6.38                                                                                           & \hspace{0.1cm}9.24   & \hspace{0.1cm}9.81    \\ \hline
\multicolumn{8}{|l|}{\hspace{3.6cm}$L_\beta$}                                                                                                                                                                                                                                                                                                                                                                                                                \\ \hline
\hspace{0.1cm}76      & \hspace{0.1cm}3.95                                                                     & \hspace{0.1cm}3.51                                                         & \hspace{0.1cm}3.70                                                                                                 & \hspace{0.1cm}1.69                                                      & \hspace{0.1cm}1.87                                                                                           & \hspace{0.1cm}2.78   & \hspace{0.1cm}2.96    \\ \hline
\hspace{0.1cm}84      & \hspace{0.1cm}5.19                                                                     & \hspace{0.1cm}4.30                                                         & \hspace{0.1cm}4.54                                                                                                & \hspace{0.1cm}2.12                                                      & \hspace{0.1cm}2.38                                                                                           & \hspace{0.1cm}3.41   & \hspace{0.1cm}3.65    \\ \hline
\hspace{0.1cm}90      & \hspace{0.1cm}6.16                                                                     & \hspace{0.1cm}4.94                                                            & \hspace{0.1cm}5.23                                                                                                & \hspace{0.1cm}2.48                                                      & \hspace{0.1cm}2.77                                                                                           & \hspace{0.1cm}3.91   & \hspace{0.1cm}4.20    \\ \hline
\hspace{0.1cm}98      & \hspace{0.1cm}6.56                                                                     & \hspace{0.1cm}5.78                                                         & \hspace{0.1cm}6.16                                                                                                & \hspace{0.1cm}2.96                                                      & \hspace{0.1cm}3.35                                                                                           & \hspace{0.1cm}4.63   & \hspace{0.1cm}5.00    \\ \hline
\hspace{0.1cm}114     & \hspace{0.1cm}9.21                                                                     & \hspace{0.1cm}7.49                                                         & \hspace{0.1cm}8.06                                                                                                & \hspace{0.1cm}3.98                                                      & \hspace{0.1cm}4.57                                                                                           & \hspace{0.1cm}6.16   & \hspace{0.1cm}6.74    \\ \hline
\multicolumn{8}{|l|}{\hspace{3.6cm}$L_\gamma$}                                                                                                                                                                                                                                                                                                                                                                                                                \\ \hline
\hspace{0.1cm}76      & \hspace{0.1cm}0.38                                                                    & \hspace{0.1cm}0.32                                                        & \hspace{0.1cm}0.34                                                                                               & \hspace{0.1cm}0.15                                                     & \hspace{0.1cm}0.18                                                                                          & \hspace{0.1cm}0.22  & \hspace{0.1cm}0.24   \\ \hline
\hspace{0.1cm}84      & \hspace{0.1cm}0.63                                                                     & \hspace{0.1cm}0.43                                                         & \hspace{0.1cm}0.46                                                                                               & \hspace{0.1cm}0.21                                                     & \hspace{0.1cm}0.24                                                                                           & \hspace{0.1cm}0.29  & \hspace{0.1cm}0.32   \\ \hline
\hspace{0.1cm}90      & \hspace{0.1cm}0.75                                                                    & \hspace{0.1cm}0.51                                                        & \hspace{0.1cm}0.55                                                                                               & \hspace{0.1cm}0.26                                                     & \hspace{0.1cm}0.30                                                                                          & \hspace{0.1cm}0.35  & \hspace{0.1cm}0.39   \\ \hline
\hspace{0.1cm}98      & \hspace{0.1cm}0.57                                                                    & \hspace{0.1cm}0.62                                                        & \hspace{0.1cm}0.67                                                                                               & \hspace{0.1cm}0.32                                                     & \hspace{0.1cm}0.37                                                                                           & \hspace{0.1cm}0.43  & \hspace{0.1cm}0.48    \\ \hline
\hspace{0.1cm}114     & \hspace{0.1cm}0.90                                                                     & \hspace{0.1cm}0.85                                                        & \hspace{0.1cm}0.92                                                                                               & \hspace{0.1cm}0.45                                                     & \hspace{0.1cm}0.52                                                                                          & \hspace{0.1cm}0.61  & \hspace{0.1cm}0.68   \\ \hline
\multicolumn{8}{|l|}{\hspace{3.6cm}$L_{Tot}$}                                                                                                                                                                                                                                                                                                                                                                                                            \\ \hline
\hspace{0.1cm}76      & \hspace{0.1cm}10.12                                                                   & \hspace{0.1cm}9.62                                                        & \hspace{0.1cm}10.02                                                                                              & \hspace{0.1cm}4.62                                                     & \hspace{0.1cm}5.00                                                                                          & \hspace{0.1cm}7.87  & \hspace{0.1cm}8.26   \\ \hline
\hspace{0.1cm}84      & \hspace{0.1cm}13.86                                                                   & \hspace{0.1cm}11.70                                                       & \hspace{0.1cm}12.20                                                                                              & \hspace{0.1cm}5.74                                                     & \hspace{0.1cm}6.28                                                                                          & \hspace{0.1cm}9.49  & \hspace{0.1cm}10.01  \\ \hline
\hspace{0.1cm}90      & \hspace{0.1cm}16.68                                                                   & \hspace{0.1cm}13.30                                                       & \hspace{0.1cm}13.9                                                                                                & \hspace{0.1cm}6.69                                                     & \hspace{0.1cm}7.28                                                                                          & \hspace{0.1cm}10.79 & \hspace{0.1cm}11.40  \\ \hline
\hspace{0.1cm}98      & \hspace{0.1cm}16.96                                                                   & \hspace{0.1cm}15.50                                                        & \hspace{0.1cm}16.32                                                                                               & \hspace{0.1cm}7.92                                                     & \hspace{0.1cm}8.79                                                                                          & \hspace{0.1cm}12.61 & \hspace{0.1cm}13.44  \\ \hline
\hspace{0.1cm}114     & \hspace{0.1cm}23.74                                                                    & \hspace{0.1cm}19.80                                                       & \hspace{0.1cm}21.04                                                                                              & \hspace{0.1cm}10.53                                                    & \hspace{0.1cm}11.78                                                                                         & \hspace{0.1cm}16.45 & \hspace{0.1cm}17.70  \\ \hline
\end{tabular}
\end{table}

%%%%%%%%%%%%%%%%%%%%%%%%%%%%%%%%%%%%%%%%%%%%%%%%%%%%
\begin{figure}[t]
\includegraphics[width=7.5cm,height=12cm]{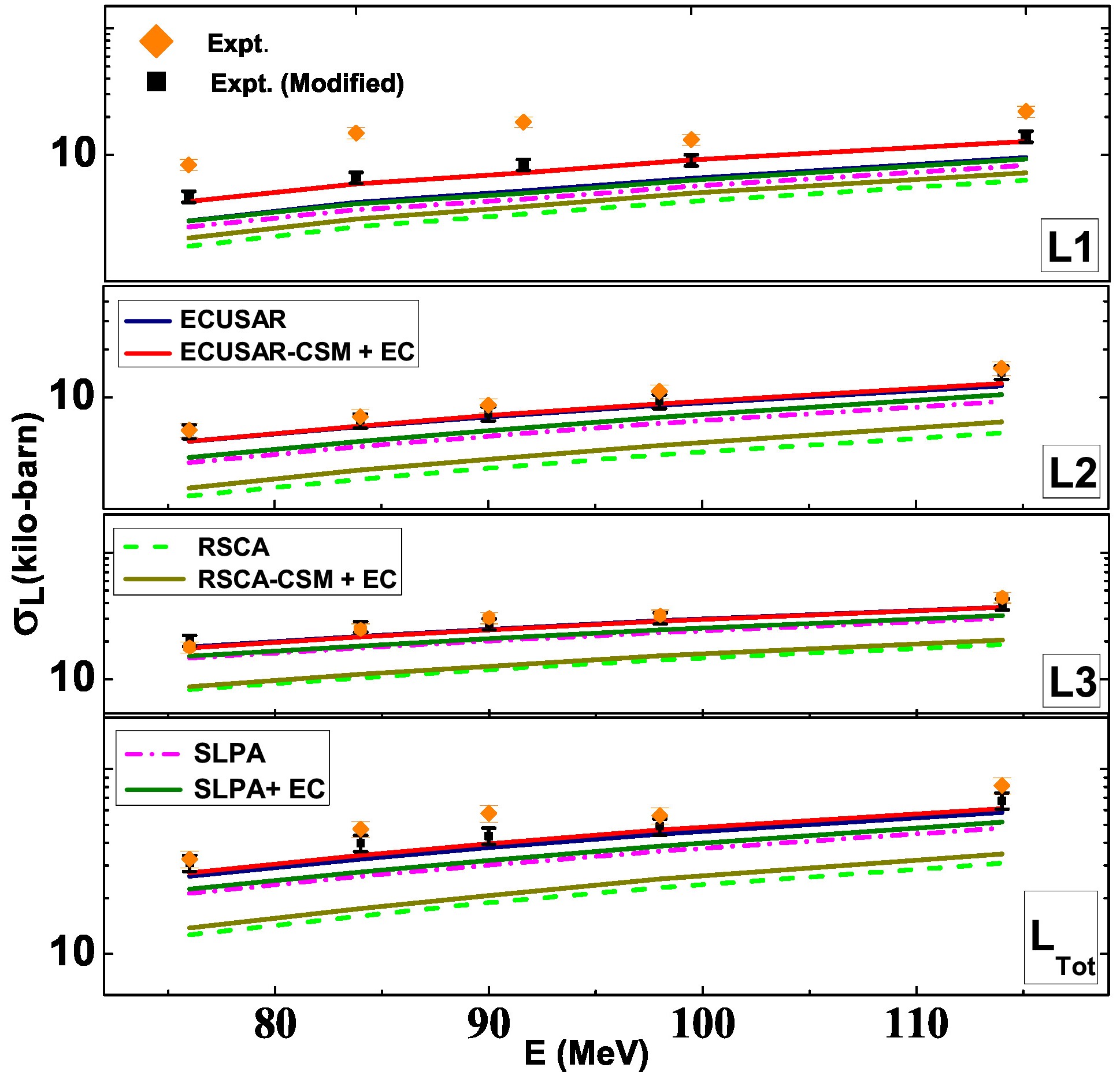}
\caption{Experimental and theoretical $L_i$-subshell $(i=1-3)$ and total $L$-shell ionization cross sections for $^{19}$F on Os collisions as a function of the impact energy. The measured data denoted by orange symbols were obtained using atomic parameters published in~\cite{campbell2003fluorescence}, those denoted by black symbols were obtained using atomic parameters which were optimized in order to have a good agreement between the theory and experiment (see text).}
\label{SIGMA_L123}
\end{figure}

\begin{figure}[t]
\centering
\includegraphics[width=6cm,height=4.5cm]{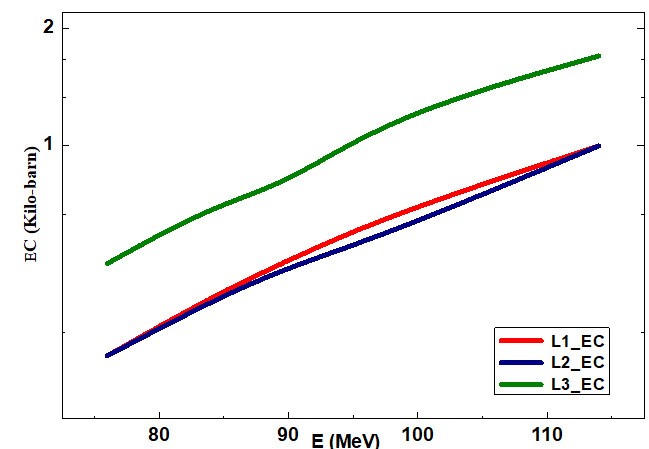}
\caption{The $L_i$-subshell $(i=1,2,3)$ ionization cross section for the electron capture from the $L$ shell of the Os target to the
$K$ shell of the $^{19}$F projectile ($L$-$K$ capture) as a function of the impact energy.}
\label{EC-GRAPH}
\end{figure}

%%%%%%%%%%%%%%%%%%%%%%%%%%%%%%%%%%%%%%%%%%%%%%%%%%%%%%%%%%%%%%%
\section{Discussions}
\label{sec:discussions}

In a recent article~\cite{singh2021evaluation}, we analyzed the major sources of errors in the measurements of $L$-subshell cross sections. We concluded that these uncertainties come from inaccurate (i) determination of target thickness, (ii) counting the number of projectile particles, (iii) background subtraction during spectrum analysis, and (iv) atomic parameters used for the conversion of (a) the measured x-ray production cross sections into $L_i$-subshell ionization cross sections and (b) the theoretical $L_i$-subshell ionization cross sections into $L$-shell x-ray production cross sections. In this work, all these aspects were considered carefully. The first source of error was managed by measuring the foil thickness by the RBS method, as shown in Fig.~\ref{RBS}. The mass thickness (target thickness in $\mu$gcm$^{-2}$) is normally measured by three techniques viz., RBS, PIXE and XRF (x-ray fluorescence). Out of these, RBS is the most accurate (see Fig.~4 of~\citet{ager2017reconsidering}). The second source of uncertainties was controlled by measuring the integrated charge count of the projectile ions in a Faraday cup placed behind the target for a fixed duration (100 s) and under two different  conditions: (a) solely the blank target frame in place and (b) the target foil in place. The ratio between the integrated charge counts for the two different conditions is  $R=\frac{nq}{nq^\prime}=\frac{q}{q^\prime}$ or $q^\prime=\frac{q}{R}$, where $n$ is the number of projectile ions of incident charge state $q$ in case of blank target, and of charge state $q^\prime$ when the target is in place. If the spectra is recorded for a long duration (say, 30 minutes), then, the total measured charge divided by $q^\prime$ will give the number of projectile ions passing through the target foil as required in Eq.~(\ref{eq:1}). The third source of error was also taken into consideration, as can be evidenced from Figs. \ref{SPECTRUM1} and \ref{SPECTRUM2}: the data points are well on the fitting profile, and the reduced $\chi$-squared values are close to 1. The fourth aspect is also addressed through a step by step evaluation, and is discussed below.

Besides the $L$-shell ionization, heavy ion collisions give rise to simultaneous ionization of the higher-shell electrons. This change of electronic environment in an atom alters the properties of the $L$ x-ray emission. As a result, the atomic parameters vary with the projectile energy as shown in Fig.~\ref{CK-W-GRAPH}. Although this effect was taken into account, the $L_i$ subshell-ionization cross sections derived from the measured $L$ x-ray production cross sections are underestimated by the ECUSAR-CSM, ECUSAR-CSM-EC, RSCA-CSM, RSCA-CSM-EC, SLPA, and SLPA-EC models, as can be seen in Fig.~\ref{SIGMA_L123}. The vacancy sharing among the $L$ subshells described by the coupled-states model (CSM) seems to have a minor role in the present collisional system. Despite the $LK$ electron capture (EC) effects are included to the various theories, the cross sections remain lower than the measured values. The EC cross sections for the $L_i$ subshells as a function of the impact energy are plotted in Fig.~\ref{EC-GRAPH}. According to the figure, the EC contribution is the largest for the $L_3$, and is almost identical for $L_1$ and $L_2$. To estimate these contributions, we used the charge-state distributions of the projectile ions inside the solid target as presented in Fig.~\ref{Q-BAR-CSD}.

The atomic parameters are used to convert the $L$ x-ray production cross section to $L_i$ subshell ionization cross section as discussed above. In general, such parameters are taken from various sources, where the authors have calculated them using different theoretical methods. In some cases, the theoretical predictions considerably deviate from each other. In particular, for $w_1$, $f_{12}$ and $f_{13}$ a spread of up to a factor of two can be observed, as shown in Table \ref{CK_W_THEORIES}. In the present work, we used the most recent theoretical parameters~\cite{campbell2003fluorescence}. 
The estimated uncertainties for $\omega_1$, $\omega_2$, $\omega_3$, $f_{12}$, $f_{13}$, and $f_{23}$ are 15\%, 5\%, 5\%, 10\%, 5\%, and 5\%, respectively. To resolve the above-mentioned discrepancy of the $L_i$ subshell ionization cross section between the measurements and the theoretical models, we varied these parameters iteratively until a good agreement was achieved; see Fig.~\ref{CK-W-GRAPH}. We present the optimized parameters in Table \ref{CK_W} and show their variation with the beam energy in Fig.~\ref{CK-W-GRAPH}. The difference between the original~\cite{campbell2003fluorescence}
values and the optimized atomic parameters is not significant, except for $\omega_1$. 

%%%%%%%%%%%%%%%%%%%%%%%%%%%%%%%%%%%%%%%%%%%%%%%%%%%%%%%%%%%%%%%%%%%%%%%%
\begin{table}
\caption{Fluorescence and CK yields for singly ionized element from different theories.}
\label{CK_W_THEORIES}
\begin{tabular}{|l|l|l|l|l|l|l|}
\hline
\multirow{2}{*}{\hspace{0.7cm}$_{76}$Os} & \multicolumn{3}{l|}{Fluorescence Yield}    & \multicolumn{3}{l|}{\hspace{0.5cm}CK Yield}                                                   \\ \cline{2-7} 
                           & $\omega_1^0$ & $\omega_2^0$ & $\omega_3^0$ & $f_{12}^0$ & \begin{tabular}[c]{@{}l@{}}$f_{13}^0$\end{tabular} & $f_{23}^0$ \\ \hline
\citet{krause1979atomic} &0.130  &0.295   &0.281   &0.16   &0.39   &0.128  \\ \hline
\citet{chen1981widths}                        & 0.088        & 0.318        & 0.282        & 0.088      & 0.636                                                 & 0.136      \\ \hline
\citet{orlic1994semiempirical}                         & 0.13         & 0.295        & 0.281        & 0.16       & 0.39                                                  & 0.128      \\ \hline
\citet{campbell2003fluorescence}                   & 0.15         & 0.318        & 0.282        & 0.07       & 0.33                                                  & 0.13       \\ \hline
\end{tabular}
\end{table}
%%%%%%%%%%%%%%%%%%%%%%%%%%%%%%%%%%%%%%%%%%%%%%%%%

%%%%%%%%%%%%%%%%%%%%%%%%%%%%%%%%%%%%%%%%%%%%%%%%
\begin{figure*}[t]
\centering
\includegraphics[width=7.5cm,height=9cm]{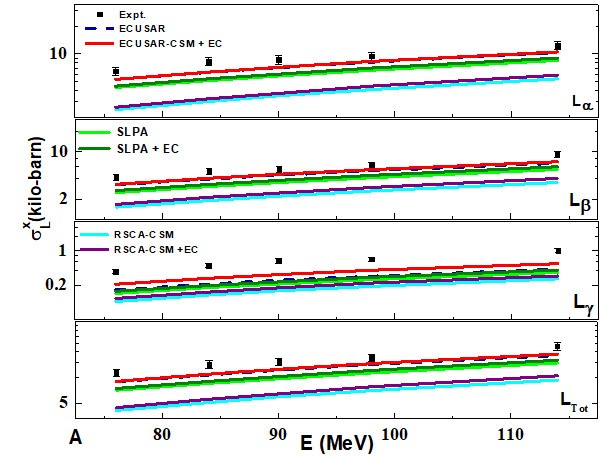}
\includegraphics[width=7.5cm,height=9cm]{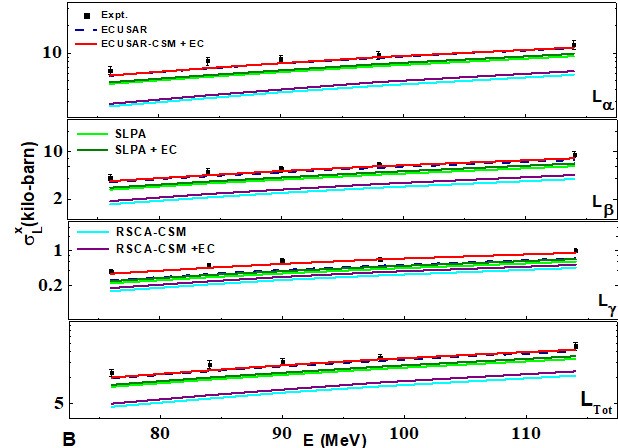}
\caption{Comparison of the measured $L_{\alpha}, L_{\beta}, L_{\gamma}$, and $L_{\mathrm{Tot}}$ x-ray production cross sections with the predictions of the different theoretical models. (A) The theoretical values were obtained using atomic parameters published in~\cite{campbell2003fluorescence}. (B) The calculations were obtained using optimized atomic parameters, which allow to have a good agreement between the theoretical and experimental $L$-subshell ionization cross sections (see text).}
\label{SIGMA}
\end{figure*}

There are two ways to study the $L_i$ subshell ionization by ion impact: by comparing the theoretical predictions (i) with the measured $L_i$ subshell ionization cross sections, and (ii) with the measured $L$ x-ray production cross sections. The former has been considered above, and the latter is examined below.

The measured x-ray production cross sections are compared with calculations in Fig.~\ref{SIGMA}. The theoretical values in Fig.~\ref{SIGMA}(A) are obtained by employing the set of atomic parameters from~\cite{campbell2003fluorescence}, while in Fig.~\ref{SIGMA}(B) the optimized set of atomic parameters are used. The various theories underestimate the measured cross sections when the original parameters are used; however, the new set of atomic parameters allows one to obtain a better agreement.

The difference between the above mentioned methods is as follows. In the first one, an individual $L_i$ subshell ionization cross section is compared, while in the second one, a combination of the contributions of the various subshells is considered. 
%Hence, in mathematical language, the first one is a differential and the second one is an integral aspect of the ionization mechanism
Hence, the mathematical expression for the former is given by a differential operation while the later constitutes an integral. From the values in Tables II and III, we notice that the agreement between the theory and experiment is better for the $L_i$ subshell ionization cross sections than for the $L$ x-ray production cross sections. Thus, the differential comparison provides a better picture that the integral one.

%%%%%%%%%%%%%%%%%%%%%%%%%%%%%%%%%%%%%%%%%%%%%%%%%%%%%%%%%%%%%%%%%%%%%%%%%%%%
\section{Conclusions}

In the present work, the $L$ x-ray production cross sections of Os were measured by 4-6 MeV/u $^{19}F^{q}$ ions of charge states $q=6+,\, 7+$ and $8+$, and compared with theoretical $L$ x-ray production cross sections. Different ionization theories such as RSCA, ECUSAR and SLPA were used for the Coulomb direct ionization. Additionally, the contribution of the $LK$ electron capture was added to each theory. The effect of multiple ionization was also considered through the modified atomic parameters. Furthermore, $L_i$ ($i=1,2,3$) subshell ionization cross sections were derived from the measured $L$ x-ray production cross sections, and compared with the corresponding theoretical counterparts. Both comparisons show the best agreement with the experiment for the ECUSAR-CSM-EC model. However, certain differences are still clearly noticed. To resolve such discrepancies, the atomic parameters were optimized to obtain a good agreement between the measurements and ECUSAR-CSM-EC model. Thus, this work gives us a convincing understanding of the $L$-subshell ionization mechanism by heavy ion bombardments, while states doubts on the atomic parameters used in the conversion of the x-ray production cross sections to the ionization cross sections. Further, our work suggests the urgent need for accurate measurements and theoretical calculations of the atomic parameters used. %in the conversion of x-ray production cross sections to ionization cross sections or vice versa.

%%%%%%%%%%%%%%%%%%%

%calculated using the $L_i$ (i=1-3) subshell ionization cross sections using ECPSSR and ECUSAR and recently recommended set of the $L_i$ (i=1-3) subshell fluorescence and CK yields with and without modifications for the multiple vacancies in the outer shells. While the measured values exhibit significant difference with the ECPSSR and ECUSAR calculations. This is particularly so when the fluorescence yields are corrected for the outer-shell multiple ionization. Although the ionization cross sections for the $^{19}F$ ions with the 6+, 7+, and 8+ charge states over the ion beam energies used in the present work are almost independent of the charge state   \cite{kumar2017shell}, the multiple ionization effect is essentially equal in the ECPSSR-MI and ECUSAR-MI calculations for q = 6+ and 7+. At q = 8+, the ECUSAR-MI agrees better with the data than the ECPSSR-MI theory.

%%%%%%%%%%%%%%%%%%%%%%%%%%%%%%%%%%%%%%%%%%%%%%%%%%%%%%

\section{Acknowledgements}

S.C. acknowledges the University of Kalyani for providing generous funding towards his fellowship. Financial support from the Science and Engineering Research Board (SERB), New Delhi (SB/FTP/PS-023/2014), to Sunil Kumar in terms of Young Scientist scheme, and the grant from IUAC, New Delhi (UF-UP-43302 project), are highly acknowledged. L. Sarkadi acknowledges the support from the Hungarian Scientific Research Fund (Grant No. K128621).
A.M.P. Mendez, D.M. Mitnik and C.C. Montanari acknowledge the financial support from the following institutions of Argentina: Consejo Nacional de Investigaciones Científicas y Técnicas (CONICET), Agencia Nacional de Promoción Científica y Tecnológica (ANPCyT), and Universidad de Buenos Aires (UBA).

The authors are grateful to the Pelletron staff for the smooth conduct of the experiments and the RBS lab for the target thickness measurements. 

%\clearpage
\bibliography{main.bbl}
\bibliographystyle{apsrev4-1}
\end{document}